\documentclass[pre,eps,aps,reprint,twocolumn,superscriptaddress]{revtex4-2}
\usepackage{graphicx}
\usepackage{amsmath}
\usepackage{amssymb}
\usepackage{color}
\usepackage[latin2]{inputenc}
\def\slaninafigdir{.}
\begin{document}
\title{%
Hydrodynamic approximations for driven dense colloidal mixtures in narrow pores
}
\author{%
Franti\v{s}ek Slanina%
}%
\affiliation{%
Institute of Physics,
 Academy of Sciences of the Czech Republic,
 Na~Slovance~2, CZ-18221~Praha,
Czech Republic%
}%
\email{
slanina@fzu.cz
}%
\author{%
Miroslav Kotrla%
}%
\affiliation{%
Institute of Physics,
 Academy of Sciences of the Czech Republic,
 Na~Slovance~2, CZ-18221~Praha,
Czech Republic%
}%
\begin{abstract}
The system of driven dense colloid mixtures is studied in one-, two-
and three-dimensional geometries. We calculate the diffusion
coefficients and mobilities for each particle type, including
cross-terms, in a hydrodynamic limit, using a
mean-field-type approximation.  The set of non-linear diffusion equations are
then solved. In one dimension, analytical results are possible. We
show that in mixtures, the ``Brazil nut'' phenomenon, or depletion of
larger particles by force of smaller ones, appears quite
generically. We calculate the ratchet current and quantify the
capability of sorting particles according to their size. We also
indicate that the ``Brazil nut'' effect lies behind the possibility of
perfect separation, where large and big particles travel in strictly
opposite direction.
\end{abstract}
%
%
\maketitle%
\section{Introduction}

Suspensions of colloid particles flowing through constrained
environments find numerous  technological applications
\cite{mat_mul_03,han_mar_08,dic_edd_hum_sto_ton_09,bur_han_mar_sch_tal_09,bec_dil_low_rei_vol_vol_16}. One
of the basic tasks in this important area is sorting the mixture of
colloid particles according to their size, shape, chirality and other
properties. Among the prominent sorting strategies let us mention the
purely hydrodynamic Segre-Silberberg effect
\cite{seg_sil_62,dic_edd_hum_sto_ton_09}, the dynamic
lateral displacement method \cite{hua_cox_aus_stu_04}, or the ratchet
mechanism
\cite{ket_rei_han_mul_00,mar_bug_tal_sil_02,reg_sch_bur_rub_rei_han_06,reg_bur_sch_rub_han_12,yan_liu_li_mar_han_zha_17,slanina_19,sla_kal_19,slanina_20}.
In   this work we focus on the latter method. It is based on the flow of
colloidal suspension through  a pore (or rather huge ensemble of pores
pierced through a membrane) under the influence of periodic external
unbiased driving \cite{mat_mul_03,mat_mul_gos_11}. The rectification of the ratchet
current is due to asymmetric geometry of the pore. Depending on the
particle size, pore diameter and frequency of the driving, the dynamics of the
particles can be dominated by Brownian motion
\cite{ket_rei_han_mul_00,reg_bur_sch_rub_han_12}, by
hydrodynamics \cite{slanina_19}, or by
combination of both \cite{sla_kal_19,slanina_20}.

Such systems are fairly well understood on one-particle
level, thus describing low-density colloidal suspensions. On the other
hand, dense colloid mixtures pose serious
problems, and many of them remain only partially resolved
\cite{wei_bec_lei_00,liu_nag_saa_wya_11,cha_cor_par_zam_12}. Here
the area touches on one side the mechanics of
granular (or in general particulate) matter \cite{ja_na_be_96a}, on another
side the geometry of sphere packgings
\cite{ast_wea_08,tor_sti_10,par_zam_10}, and on yet another
side the stochastic dynamics of exclusion processes
\cite{derrida_98,kri_kru_10}.

In this work, {we consider particles moving along a tube which is not
plain and straight  but is shaped and structured in various
ways. The particles can be grains of sand of size around one
millimeter falling down by gravity or electrostatically charged colloid particles of
micrometer size driven by external electric field or blood cells
flowing in a vein or something analogous.
The particles
scatter and interact among each other. They may move freely or get
jammed. They may concentrate in certain places and leave other
places nearly void. Here we aim at describing such a complex swirl in
a schematic and highly simplified way. The buildup of a schematic
description proceeds in several steps. In the first step we replace
the Newtonian mechanics of the particles accompanied by energy
dissipation and interaction with surrounding fluid (be it air or
water or oil or anything else) by purely stochastic process. We also replace the
irregular shapes of the particles by perfect spheres.  Elastoplastic
properties of the particles as well as fluid-mediated hydrodynamic
interactions between them are replaced by simple requirement that
the centers of the spheres may not come closer than the sum of their
radiuses.
In the second step we discretize the Euclidean
space, dividing it into cells of finite size. Specific position of
particles within a single cell is neglected and we retain just
information on which cell contains which particle. Particles hop
between cells stochastically. In the third step we take a look at the
system from a distance. We cannot distinguish each single cell any
more but they look like a continuum. In this view the dynamics
becomes deterministic again and is described by (non-linear)
partial differential equations. All non-trivial features of the
particles we started with, i. e. their size, weight, mobility etc.,
are embodied in the non-linearities of these equations.  The only
task which remains now is to solve the equations, either
numerically or, in a lucky case, analytically. The aim of this work
is to provide a host of such solutions.}

We shall suppose that the movement of the colloid
particles is purely stochastic, reducing the hydrodynamic effects to
bare homogeneous external drift, or constant bias in hopping
rates. This is a radical trivialization of the problem, as the
hydrodynamics of the fluid medium surrounding all the particles not
only provides a driving force, but also substantially changes the
inter-particle interactions, providing a long-range and fundamentally
non-additive contribution to particle-particle and
particle-wall interaction. Treating these effects is a separate and
difficult task, therefore we shall skip it completely in this work.

Therefore, we base our models on the ground of stochastic exclusion
processes. The prominent representative of this set of models is the
asymmetric simple exclusion process (ASEP), which was solved exactly by
several methods
\cite{gwa_spo_92b,do_do_mu_92,schutz_93a,de_ev_pa_93,de_ja_le_spe_93a,sandow_94,sch_zia_95,schutz_97,schutz_01,bly_eva_07}. The
most
stunning feature of this model is the absence
of spatial correlations in stationary state. It was shown
that this feature is moderately robust, in the sense that various
generalizations of ASEP suggested in the literature
\cite{gab_kra_red_10,bas_moh_09,bas_moh_10,bas_moh_10a,pri_ayy_jai_14,pin_gov_13,krapivsky_13}
on one
side do exhibit spatial correlations, but on the other side these
correlations are short-ranged and in a few special cases can be
calculated exactly by a cluster mean-field approximation
\cite{gab_kra_red_10,bas_moh_09,pin_gov_13}. In these cases the
correlations decay with a strictly
exponential tail.

In a recent series of papers
\cite{hum_kot_net_sla_20,hum_kot_sla_21,sla_kot_net_22} we suggested
another
generalization of ASEP, following earlier works of Refs
\cite{kip_lan_oll_94,seppalainen_99,kip_lan_99,bec_nel_cle_par_bro_13,bec_nel_cle_par_bro_14,ari_kra_mal_14,ari_kra_mal_17},
where the generalized symmetric exclusion process was
investigated. The core of the generalization consists in
allowing not just one particle on a lattice point, but a number not
exceeding a fixed limit $k$. Moreover, we allowed several species of
particles characterized by different
sizes. In this case the exclusion constraint means that the sum of
sizes of all particles on a single site
cannot exceed the limit $k$. Most importantly, we found that in such
generalized ASEP model the long-range
correlations in stationary state decay algebraically, with an
exponent whose value is conjectured to be exactly $2$. Despite the
long-range correlations, the mean-field and cluster mean-field
approximations were found to be reasonably reliable in calculation of
nearest-neighbor correlations as well as particle currents. Indeed,
currents depend explicitly only on short-range correlations and the
non-trivial long-range correlations enter just indirectly.

This fact encourages us to perform a continuous
approximation of the generalized ASEP model. The coarse-graining
procedure with proper scaling limit leads to hydrodynamic equations in
which only long-wavelength modes associated with locally conserved
quantities survive, while all the fast-decaying short-wavelength modes
are projected out \cite{spohn_91}. To establish the validity of the
hydrodynamic limit is a
difficult task, but there are cases where it is resolved with
mathematical rigor
\cite{spohn_91,kip_lan_oll_94,seppalainen_99}. There are various
sources of
difficulties. For example, the long-range
stationary correlations may easily spoil the scaling limit. We found
that
such correlations are indeed present in the generalized ASEP model
studied here
\cite{sla_kot_net_22}. Fortunately, for the one-component variant of
our generalized ASEP, with slightly different hopping rates, the
existence of hydrodynamic limit was proved
\cite{seppalainen_99}, which encourages us to assume that the
hydrodynamic limit is well defined also in our case, and especially
for systems with
several types of particles.

Hydrodynamic equations are just expressions of
conservation laws. For each conserved quantity the temporal change of
its local density equates to the divergence of the associated
current. The form of the equations may differ depending on the number
and type (scalar, vector, etc) of the conserved quantities. In this
article it will be the set of a few scalars, namely the densities of
particles of several types.
The currents are themselves functionals of all the densities. These
are the constitutive relations characteristic of the model in
question. In the
simplest case such functional is local and depends on derivatives of
the densities up to a finite order.

At this point we must stress that the coarse-graining and continuum
limit can be performed in (at least) two different scaling regimes
\cite{kip_lan_99}. In
the Eulerian scaling, time is scaled linearly with space, while in
diffusion scaling time scales with square of space. In diffusion
scaling we assume that in the current functional the first spatial
derivative is connected to the diffusion coefficient,
while higher derivatives are irrelevant in renormalization-group
sense. On the other hand, in Eulerian
scaling we assume that already the first spatial
derivative  is irrelevant and the dynamics is ballistic. To study the
fluctuations, diffusion term is added phenomenologically, together
with Gaussian noise, whose properties are related to the diffusion
coefficient via fluctuation-dissipation theorem. This way we arrive at
fluctuating hydrodynamics
\cite{tre_ara_sig_81,spohn_14,spo_sto_15,pop_sch_sch_14,pop_sch_sch_sch_15,pop_sch_sch_sch_16,che_gie_hik_sas_18}.
Alternative route is taken in the macroscopic fluctuation theory
\cite{ber_des_gab_jon_lan_02,ber_des_gab_jon_lan_15} which is based on
assumption that not only the microscopic dynamics of the system, but
also its adjoint dynamics possess well-defined hydrodynamic
limit. (This is trivially satisfied for reversible dynamics, as it is
self-adjoint.)

In our work we shall follow the
path of diffusion scaling. Then, the
basic step in solving the model in question is to establish the
density dependence of the transport coefficients. Let us stress once
more, that in principle the long-range correlations may lead to
non-vanishing higher spatial derivatives or non-locality, but we
proceed assuming these effects irrelevant.

Calculation of diffusion and mobility coefficients is simplified in
models with gradient property
\cite{spohn_91}.  For non-gradient systems with reversible dynamics,
variational procedure is in principle exact \cite{spohn_91} but
computationally demanding
\cite{ari_kra_mal_17,ari_kra_mal_18}. Unfortunately, the models
investigated in this work
are not gradient and not reversible, so we have to resort to
approximations. In a low-density limit it is possible to use
perturbative approach \cite{bru_cha_12}, but here we need a solution
for entire range of allowed densities.
Mean-filed-type approximations often
lead to results which deviate very little from exact numbers
\cite{bec_nel_cle_par_bro_14,ari_kra_mal_14,teo_sho_17,sellito_20}. However,
we should be
careful in interpreting the results. For example, when investigating
subtle effects, like Casimir forces
\cite{cat_bri_mar_nor_sot_06,kir_ort_sen_13,ami_kaf_kar_15}, such tiny
deviations
translate in substantial differences \cite{sellito_20,sellito_20a}. On
the other hand, the argument based on variational principle
\cite{spohn_91} suggests that the mean-filed-type approximations give
weak but exact upper bound on the true diffusion coefficient
\cite{ari_kra_mal_17,ari_kra_mal_18}.

The aim of our work is calculation of transport coefficients for
generalized ASEP model with several types of particles differentiated
by their size and with cell capacity $k$, $1<k<\infty$. The
calculation relies on a mean-field-type approximation, neglecting
spatial correlations in a non-equilibrium state. We then apply the
obtained formulas to specific examples of driven colloid mixtures in quasi
one-dimensional narrow pores. When mapped on piecewise one-dimensional
geometries, it is possible to obtain analytical results. One of the
specific questions investigated will be the possibility of perfect
separation of particles according to their size.

\section{From continuous to discrete model and back}

\subsection{Local mixing approximation}

In our previous work \cite{sla_kot_net_22} we studied the system of
dense colloid mixture using an approach we called local mixing
approximation. It consists in emulating the stochastic motion of
hard spherical particles by a discrete generalized asymmetric
exclusion (ASEP) process. Let us briefly recall the idea here. For more
details, we refer the reader to Ref.  \cite{sla_kot_net_22}. For
simplicity, the
formulas written in this
subsection will describe movement in one dimension only, but we shall
consider general dimension in the following.

We consider the system of spherical colloid particles interacting
by steric repulsion. The particles may be of different sizes, the
diameter of $i$-th particle being $d_i$.
In absence of other particles, each particle  taken individually
would perform a Brownian motion with bias. For
the coordinate $x_{i}$ of the center of $i$-th particle we have
a standard stochastic equation
\begin{equation}
\mathrm{d} x_{ i}(t)=f_i\mathrm{d} t
+\mathrm{d} W_{ i}(t)
\label{eq:stochcontinuous}
\end{equation}
where $W_{ i}(t)$ is ensemble of independent Wiener
processes,
$(\mathrm{d}W_{ i})^2=2D_i\mathrm{d} t$.
The diffusion
coefficient $D_i$ and the drift $f_i$ depend only on the
diameter of the particle (they are both inversely proportional to
$d_i$). In the following we shall classify the particles into $M$
types according to their diameter. Then, the transport coefficients
will be denoted $D_{0\alpha}$ and $f_{0\alpha}$ for all particles
which belong to type $\alpha$. The ``$0$'' in the index indicates,
that the diffusion coefficient and drift pertain to non-interacting
Brownian motion.

The hard-sphere interaction between particles is expressed by the
constraint
\begin{equation}
|x_{ i}(t)-x_{ j}(t)|> \frac{1}{2}(d_{i }+d_{j })
\label{eq:stericcontinuous}
\end{equation}
which must be valid at all times. The trajectories produced by the
process (\ref{eq:stochcontinuous}) but violating
(\ref{eq:stericcontinuous}) are forbidden. This constraint makes the
dynamics non-trivial.
{We may describe it formally as follows. Let us suppose we find
the probability measure $\mu_0[\{x_i(t)\}_{i=0}^N]$ in the space of
all possible trajectories $x_i(t)$, $i=1,\ldots,N$, $t\in [t_0,t_1]$ of the ensemble
if $N$ particles, which started at time $t_0$ and ended at time
$t_1$.
The measure $\mu_0$ is generated by the stochastic
process (\ref{eq:stochcontinuous}). Then, in this space of
trajectories we define the indicator
function $\chi[\{x_i(t)\}_{i=0}^N]$ such that $\chi=1$ if
(\ref{eq:stericcontinuous}) is satisfied for all $i,j$ and all $t\in   [t_0,t_1]$. Otherwise, $\chi=0$.
Then the proper measure of our process in the space of all
trajectories is
\begin{equation}
\mu[\{x_i(t)\}_{i=0}^N] = \frac{1}{\Xi}\mu_0[\{x_i(t)\}_{i=0}^N]\chi[\{x_i(t)\}_{i=0}^N]
\end{equation}
where the number $\Xi$ ensures the proper normalization.
}

{We performed practical computer simulation of this
process in our previous work \cite{sla_kot_net_22}. Interested reader
may find the details on the numerical implementation there.}

To simplify the situation, we emulate the continuous process
(\ref{eq:stochcontinuous})
by a discrete one. To
this end, we partition the space into disjoint cells and neglect the
dynamics of the particles within the cells.
The particles can hop from one cell to the next one with rates
depending on the particle type. The position of $i$-th particle
evolves according to
\begin{equation}
x_{ i}(t) - x_{ i}(0) = S_+(t)-S_-(t)
\label{eq:stochdiscrete}
\end{equation}
where $S_+(t)$ and $S_-(t)$ are Poisson processes with rates
$a_\alpha$ and $b_\alpha$, respectively. The rates depend only in the type $\alpha$
of the particle $i$.
They are related to the
properties of the process  (\ref{eq:stochcontinuous}) as
$a_\alpha-b_\alpha=f_{0\alpha}$, $a_\alpha+b_\alpha=2D_{0\alpha}$.
For simplicity, we fix the unit length as the cell size.

The constraint
(\ref{eq:stericcontinuous}) is taken into account by the requirement
that only certain configurations of particles can enter into the
cell. More specifically, we fix a cell capacity $k$ and weight factors
$c_\alpha$ which are related to particle diameters. If there are
$n_\alpha$ particles of type $\alpha$ inside a cell, we require that
the cell capacity is not exceeded, i.e.
\begin{equation}
\sum_{\alpha=1}^M c_\alpha n_\alpha \leq k \;.
\label{eq:stericdiscrete}
\end{equation}
This constraint must be satisfied at all cells and all times.
Trajectories produced by the process (\ref{eq:stochdiscrete}) but
violating the constraint (\ref{eq:stericdiscrete}) are forbidden.

Such an approximation effectively assumes that the dynamics within
cell is fast, so that the particles are mixed on the level of cells
and for description of the global behavior of the mixture it is
sufficient to consider inter-cell hopping constrained by the condition
(\ref{eq:stericdiscrete}). Hence the name local mixing
approximation.

{In our previous work \cite{sla_kot_net_22} we compared
the simulation results coming from the continuous stochastic process
(\ref{eq:stochcontinuous}),  (\ref{eq:stericcontinuous})
with those of the
discrete process (\ref{eq:stochdiscrete}),
(\ref{eq:stericdiscrete}). This comparison indicates how the
parameters $k$ and $c_\alpha$ should be chosen in order to get optimal
match of the continuous and discrete processes.}

\subsection{Hydrodynamic approximation}

Hydrodynamic approximation assumes that relative changes of particle
concentrations are small on the scale of lattice constant. It is a
non-trivial problem of mathematical physics, whether a
well-defined hydrodynamic limit exists \cite{spohn_91}. Intuitively,
we expect that dynamics of slow modes related to conserved quantities
(in our case densities of particles of each type)
give rise to hydrodynamic equations of diffusion type, while all other
fast modes are effectively averaged out. There is quite important set of
models in which the existence of hydrodynamic limit is proved
rigorously  \cite{spohn_91}. However, here we remain on rather
heuristic level. The task is to calculate density dependence of
transport coefficients, i. e. the diffusion coefficients and drifts. In our
derivation we rely on
Refs. \cite{bec_nel_cle_par_bro_14,ari_kra_mal_14,teo_sho_17,sellito_20},
but we generalize these results for the case of multicomponent
system. More detailed derivation of our formulas using an alternative
method will be published elsewhere \cite{net_sla_prep}.

In the following we suppose there are just two types of particles,
called ``small'' and
``big'', with size factors $c_s=1$, $c_b=2$.
Therefore, the number of $n_s$ of small  and $n_b$ of big particles in a
single cell must satisfy the constraint $n_s+2n_b \le k$. The key
quantities in the derivation of transport coefficients are the
probabilities
\begin{equation}
\begin{split}
\overline{P}_s=\,&\mathrm{Prob}\{n_s+2n_b=k\}\\
\overline{P}_b=\,&\mathrm{Prob}\{n_s+2n_b \ge k-1\}
\end{split}
\end{equation}
of such configurations that do not accommodate any extra small, and
big particle, respectively. In a homogeneous stationary state these
probabilities are functions of average densities $\rho_s$ and $\rho_b$
of small and big particles. In hydrodynamic limit, densities are
(slowly changing) functions of coordinate, and so are also
$\overline{P}_s(\rho_s,\rho_b)$ and
$\overline{P}_b(\rho_s,\rho_b)$. The two-component
diffusion problem is governed by equations
\begin{equation}
\begin{split}
\frac{\partial}{\partial t}\rho_s=
\nabla\cdot\Bigg(
D_{SS}(\rho_s,\rho_b)\nabla\rho_s\\+
D_{SB}(\rho_s,\rho_b)\nabla\rho_b
-{\bf f}_s(\rho_s,\rho_b)\rho_s\Bigg)\\
\frac{\partial}{\partial t}\rho_b=
\nabla\cdot\Bigg(
D_{BS}(\rho_s,\rho_b)\nabla\rho_s\\+
D_{BB}(\rho_s,\rho_b)\nabla\rho_b
-{\bf f}_b(\rho_s,\rho_b)\rho_b\Bigg)
\label{eq:difeqSB}
\end{split}
\end{equation}
and the transport coefficients can be expressed, using methods of
\cite{bec_nel_cle_par_bro_14,ari_kra_mal_14}, as
\begin{equation}
\begin{split}
{\bf f}_s =\,& {\bf f}_{0s}(1-\overline{P}_s(\rho_s,\rho_b))\\
{\bf f}_b =\,& {\bf f}_{0b}(1-\overline{P}_b(\rho_s,\rho_b))\\
D_{SS} =\,& D_{0s}\Big(1-\overline{P}_s(\rho_s,\rho_b)
+\rho_s\frac{\partial}{\partial\rho_s}\overline{P}_s(\rho_s,\rho_b)\Big)\\
D_{SB} =\,& D_{0s}
\rho_s\frac{\partial}{\partial\rho_b}\overline{P}_s(\rho_s,\rho_b)\\
D_{BS} =\,& D_{0b}
\rho_b\frac{\partial}{\partial\rho_s}\overline{P}_b(\rho_s,\rho_b)\\
D_{BB} =\,& D_{0b}\Big(1-\overline{P}_b(\rho_s,\rho_b)
+\rho_b\frac{\partial}{\partial\rho_b}\overline{P}_b(\rho_s,\rho_b)\Big)\;.
\label{eq:Dandf}
\end{split}
\end{equation}
We denoted $D_{0s}$ and $D_{0b}$ are diffusion coefficients of pure
system of small and big particles at infinite dilution, and similarly
${\bf f}_{0s}$ and ${\bf f}_{0b}$ drift of small and big particles at
infinite dilution.

Besides the long-wavelength hydrodynamic limit we apply also
mean-field approximation, neglecting correlations in occupation of
neighbor sites.
{We addressed this question in our earlier work
\cite{hum_kot_net_sla_20}, where we investigated the generalized ASEP
model with several types of particles. It was proved there, that in the mean-field
approximation, the probabilities of one-site configurations follow
truncated Poisson distribution in the case of one type of particles,
double truncated Poisson distribution in the case of two types of
particles,
etc. Therefore, we apply the multiple truncated Poisson distribution
as the essence of the mean-filed approximation we make.}

{Therefore, in this approximation,}
the probability of having $n_s$ small and $n_b$ big
particles in one cell is given by double truncated Poisson distribution
\begin{equation}
P(n_s,n_b)=\frac{\lambda_s^{n_s}\lambda_b^{n_b}}{Z n_s! n_b!}
\label{eq:doublepoisson}
\end{equation}
where
\begin{equation}
Z(\lambda_s,\lambda_b)=\sum_{n_s=0}^k
\sum_{n_b=0}^{[(k-n_s)/2]}
\frac{\lambda_s^{n_s}\lambda_b^{n_b}}{ n_s! n_b!}
\label{eq:doublepoissonZ}
\end{equation}
is the partition function.
{The two parameters  $\lambda_s$ and $\lambda_b$
of
the double Poisson distribution, which belong to the
small and big particles, respectively, fix the average
densities. Borrowing a term from equilibrium statistical physics, we shall call the parameters
$\lambda_s$ and $\lambda_b$ fugacities, keeping in mind that they are
just numbers that parametrize the probability distribution.}

{There is a one-to one correspondence between the pair
of fugacities $\lambda_s$ and $\lambda_b$ and the pair of particle
densities $\rho_s$ and $\rho_b$. Indeed, the fugacities determine
the probability distribution of on-site particle configurations and this in turn
determines the average densities.}
Therefore, we suppose that the dynamics in hydrodynamic limit can
be formulated using time-and-position-dependent fugacities   $\lambda_s(x,t)$ and
$\lambda_b(x,t)$ instead of time-and-position-dependent densities  $\rho_s(x,t)$ and
$\rho_b(x,t)$. These two formulations are translated one to the other
by the functions which express stationary and homogeneous densities
through fugacities
\begin{equation}
\begin{split}
\rho_s=\,&R_s(\lambda_s,\lambda_b)\equiv
\lambda_s\frac{\partial \ln Z(\lambda_s,\lambda_b)}{\partial\lambda_s}\\
\rho_b=\,&R_b(\lambda_s,\lambda_b)\equiv
\lambda_b\frac{\partial \ln
Z(\lambda_s,\lambda_b)}{\partial\lambda_b}\;.
\label{eq:Rfunctions}
\end{split}
\end{equation}
{These formulas follow directly from the probabilities
given by (\ref{eq:doublepoisson}) and (\ref{eq:doublepoissonZ}).}

In stationary state it leads to a great simplification.
Indeed, as can be checked by insertion of (\ref{eq:Rfunctions}) and
(\ref{eq:Dandf}) into
diffusion equations (\ref{eq:difeqSB}) with $\partial\rho_s/\partial
t=\partial\rho_b/\partial t=0$, the stationary state satisfies
the equations for fugacities
\begin{equation}
\begin{split}
\nabla\cdot
\Big[\Big(\frac{R^2_\alpha(\lambda_s,\lambda_b)}{\lambda_\alpha(x)}\Big)
\big(D_{0\alpha}\frac{\nabla\lambda_\alpha(x)}{\lambda_\alpha(x)}-{\bf
f}_{0\alpha} \big)\Big]=0\;.
\label{eq:stactwotypes}
\end{split}
\end{equation}
The index $\alpha\in\{S,B\}$ denotes the particle type.
The equations (\ref{eq:stactwotypes}) are still coupled through the functions
$R^2_\alpha(\lambda_s,\lambda_b)$, but do
not contain cross-diffusion terms. This fact serves us as a basis for
further calculations.

{In order to avoid confusion, we also stress that at
the same time as the hydrodynamic approximation is made, the
particles become effectively point-like. Therefore, no effects can occur
stemming from the incommensurability of particle size and spatial
period of the geometry, as it was observed in \cite{der_vic_95} and also in
our earlier work \cite{slanina_09b}.
}

\section{Toy examples: piecewise one-dimensional geometry}

\subsection{Simplifications in 1D}

On a one-dimensional segment, the stationary currents of all types of
particles are constants independent of position. This makes the
one-dimensional geometry fundamentally
different from higher dimensions. The diffusion equations formulated
for fugacities (\ref{eq:stactwotypes}) can be written as
\begin{equation}
\begin{split}
&j_\alpha\frac{\lambda_\alpha(x)}{R_\alpha^2(\lambda_s(x),\lambda_b(x))}=-D_{0\alpha}
\frac{\partial\lambda_\alpha(x)}{\lambda_\alpha(x)\partial
x}
+f_{0\alpha}\;.
\label{eq:difeq1dforlambda}
\end{split}
\end{equation}
where $j_\alpha$ is the constant current of the particles of type
$\alpha$ along the segment. The equations (\ref{eq:difeq1dforlambda})
are simplified with respect to (\ref{eq:stactwotypes})
in the sense that they are of first order in spatial
derivative. However, a serious complication remains here, namely the
fact that the equations mix the
dependence on all particle types through the functions
$R_\alpha^2(\lambda_s(x),\lambda_b(x))$ which contain fugacities of all
particle types as arguments. This complication is absent in two
special situations we shall study in the following two subsections.

\subsection{One type of particles in sawtooth potential}

The equations (\ref{eq:difeq1dforlambda}) are particularly simple, if we
allow just one type of particles. At the same time, the drift can be
spatially dependent. Thus, we have the equation
\begin{equation}
D_0\frac{\partial\lambda(x)}{\lambda(x)\partial x}=f_0(x)-
j\frac{\lambda(x)}{R^2(\lambda(x))}\;.
\label{eq:forlambda1dsawtooth}
\end{equation}
We suppose the drift term has the form $f_0(x)=F-\frac{\partial V(x)}{\partial x}$
where $F$ is constant external force and $V(x)$ is a periodic
potential. As a simplest choice we use the sawtooth form
\begin{equation}
\begin{split}
&V(x) = \frac{v}{L_+}x\qquad \mathrm{for}\;\; x\in (0,L_+)\\
&V(x) = -\frac{v}{L_-}x\qquad \mathrm{for}\;\; x\in (-L_-,0)\\
&V(x+L_++L_-)=V(x)\qquad\forall x\;.
\end{split}
\label{eq:sawtooth}
\end{equation}
In this case the drift is a piecewise constant
function and this enables us to find the solution in terms of closed analytic
formulas.
The differential equation (\ref{eq:forlambda1dsawtooth}) together with the periodic
condition $\lambda(-L_-)=\lambda(L_+)$ yields the pair of
transcendental equations for the two fugacities
$\lambda_0\equiv\lambda(0)$ and $\lambda_1\equiv\lambda(L_+)$
\begin{equation}
\begin{split}
\int_{\lambda_1}^{\lambda_0}\frac{R^2(\lambda)d\lambda}{
(\frac{v}{L_+}-F)\lambda R^2(\lambda)+j\lambda^2}
=\frac{L_+}{D_0}\\
\int_{\lambda_1}^{\lambda_0}\frac{R^2(\lambda)d\lambda}{
(\frac{v}{L_-}+F)\lambda R^2(\lambda)-j\lambda^2}
=\frac{L_-}{D_0}\;.
\end{split}
\end{equation}
When solved, we obtain the average particle density
$\overline{\rho}=\int_{-L_-}^{L_+}\rho(x)dx/(L_-+L_+)$ as
\begin{equation}
\begin{split}
\overline{\rho}=\frac{D_0}{L_-+L_+}\int_{\lambda_1}^{\lambda_0}\Bigg[
\frac{1}{
(\frac{v}{L_+}-F)\lambda R^2(\lambda)+j\lambda^2}
+\\
+\frac{1}{
(\frac{v}{L_-}+F)\lambda R^2(\lambda)-j\lambda^2}
\Bigg]R^3(\lambda)d\lambda
\end{split}
\end{equation}

\begin{figure}[t]
\includegraphics[scale=0.85]{%
\slaninafigdir/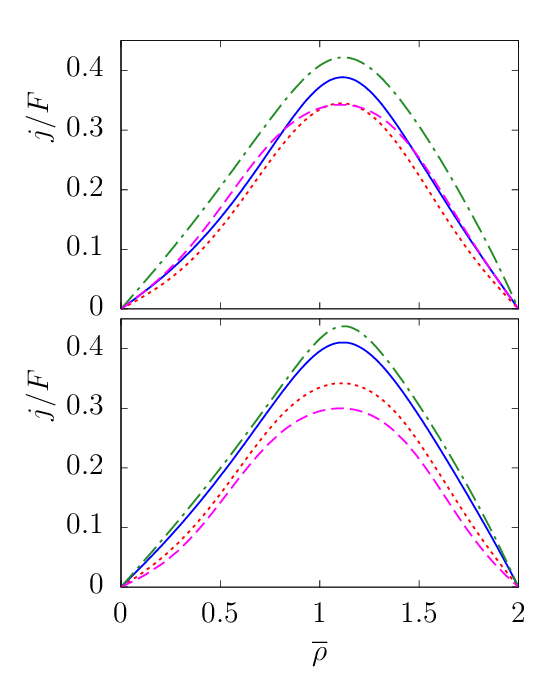}
\caption{
{
Density dependence on the particle current in one-dimensional geometry with sawtooth
potential (\ref{eq:sawtooth}) with parameters $k=2$, $D_0=0.1$,
$v=0.5$.
In the upper panel we use shape parameters $L_+=0.7$,
$L_-=0.3$ and force $F=0.3$ (solid line), $F=-0.3$ (dotted line),
$F=0.6$ (dot-dashed line), and  $F=-0.6$ (dashed line).
In the lower panel we use force  parameters
$F=0.5$ (solid line, dot-dashed
line) and $F=-0.5$ (dashed line, dotted line) and shape
$L_+=0.7$,
$L_-=0.3$ (solid line, dotted line) and
$L_+=0.9$,
$L_-=0.1$ (dot-dashed line, dashed line).
}
}
\label{j-1d-k2-D0p1-v0p5-lpm0p3and0p7-f0p3}
\end{figure}

We calculated the current-density diagram for the simplest case $k=2$
where the function determining density in terms of fugacity is

\begin{equation}
R(\lambda) = \frac{\lambda+\lambda^2}{1+\lambda+\frac{1}{2}\lambda^2}\;.
\end{equation}

We show a typical behavior in
Fig. \ref{j-1d-k2-D0p1-v0p5-lpm0p3and0p7-f0p3}.
{As expected, the
density dependence of the current has a maximum. Although the detailed
form of the current-density curve depends on parameters of the model
like the force $F$ and geometry determined by $L_+$, $L_-$, the
position of the maximum remains nearly unchanged. We also observe, that} for average
density approaching either $\overline{\rho}\to 0$ or
$\overline{\rho}\to 2$ the dependence is linear, because it
corresponds to the regimes of nearly independent particles and nearly
independent holes, respectively. However, when the density increases
from zero, we observe enhancement of the current over the linear
asymptotics. Such interaction-induced enhancement can be easily
understood if we realize that free particles become partially trapped
around the minimum of the potential $V(x)$, which leads to decrease in
current. Such trapping is less severe for particles interacting by
exclusion, because the effective capacity of the trap is limited by
the exclusion principle. Indeed, for $k=2$ we must have $\rho(x)\le 2$
everywhere, including the neighborhood of the minimum of the
potential, while for free particles $\rho(x)$ can be arbitrarily
large. Less particles trapped means more current, hence the
super-linear increase of current for small densities. Analogous
behavior is also observed close to the maximum density (which is $2$
in this case), due to trapping of interacting holes. Although there is
no particle-hole symmetry in this model, holes behave in a
qualitatively similar manner as particles.

\begin{figure}[t]
\includegraphics[scale=0.85]{%
\slaninafigdir/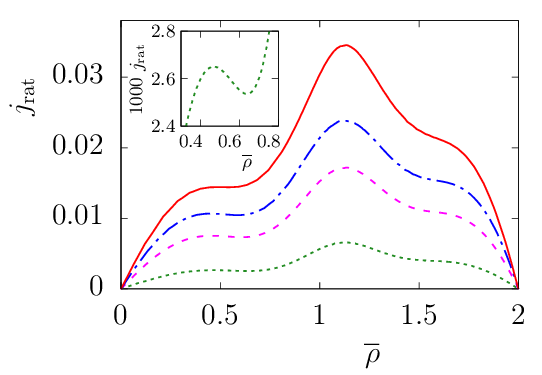}
\caption{Density dependence on the ratchet current in one-dimensional geometry with sawtooth
potential (\ref{eq:sawtooth}). The diffusion coefficient is
$D_0=0.1$ and the geometry parameter $v=0.5$. The other parameters are $L_+=0.7$,
$L_-=0.3$, and $F=0.3$ (dotted line),
$L_+=0.7$,
$L_-=0.3$, and $F=0.5$ (dashed line),
$L_+=0.7$,
$L_-=0.3$, and $F=0.6$ (dash-dotted line),
$L_+=0.9$,
$L_-=0.1$, and $F=0.5$ (solid line).
In the inset, detail of the same data, for  $L_+=0.7$,
$L_-=0.3$, and $F=0.3$.
}
\label{ratchet-1d-k2-D0p1-v0p5}
\end{figure}

For $L_+>L_-$ the ``easy'' direction of the flow under the influence
of external driving $F$ is positive, i.e. rightward. From
Fig. \ref{j-1d-k2-D0p1-v0p5-lpm0p3and0p7-f0p3} we can see that such
intuitive reasoning holds true not just for nearly-free particles, but at
all densities. We observed, by varying the parameters of geometry as
well as driving and diffusion coefficient, that this conclusion is
generic. If we switch the external driving regularly between the
values $|F|$ and $-|F|$, the current averaged over many periods will
exhibit the ratchet effect. For very slow switching the ratchet
current is just the combination of stationary currents
\begin{equation}
j_\mathrm{rat}=\frac{1}{2}\Big(j_{F=|F|}+j_{F=-|F|}\Big)\;.
\end{equation}
We show in Fig. \ref{ratchet-1d-k2-D0p1-v0p5} the ratchet current for
several sets of the parameters of the model. As we already mentioned,
the current is always larger in the ``easy'' direction, independently
of the average density, so there is no current reversal at higher
densities, contrary to the geometries we shall study later in this
paper. Interesting feature, which we observed quite generically with
the sawtooth potential, is
the presence of two inflection points at intermediate densities. This
is related to the already mentioned fact that the current behaves
super-linearly both near the zero and near the maximum density. For
small enough driving $F$, the effect may even lead to weak
non-monotonicity of the ratchet current, as shown in the inset of Fig.
\ref{ratchet-1d-k2-D0p1-v0p5}.

\subsection{Two types of particles in pocket geometry}

\begin{figure}[t]
\includegraphics[scale=0.4]{%
\slaninafigdir/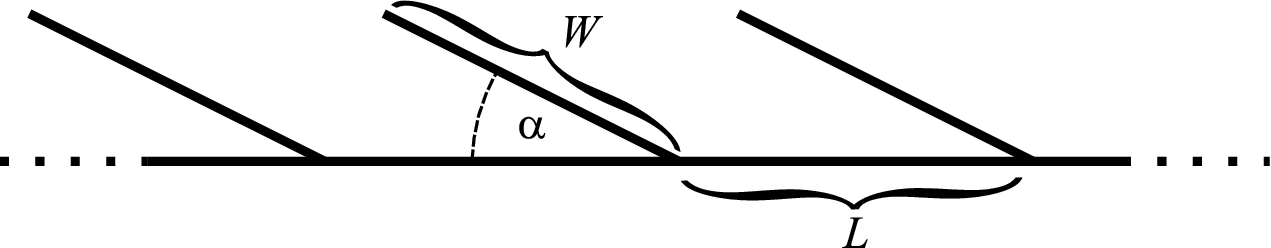}
\caption{Scheme of the pocket geometry. On a straight line (backbone)
at regular distances $L$, segments of length $W$ are attached
(pockets). The pockets are tilted at angle $\alpha$.
}
\label{scheme-pockets}
\end{figure}

Another easily soluble case corresponds to particles in the pocket
geometry sketched in Fig \ref{scheme-pockets}.
On a straight line (we shall call it backbone), at regularly spaced
points at distance $L$ of each other, tilted segments of length $W$
are attached. These segments represent dead ends, so that no current
can flow through them in stationary state, but they act as traps accumulating certain
portion of particles, thus reducing the current along the backbone. We
shall call these segments pockets.

The pockets are tilted to the left at angle $\alpha$. This has the effect that if
the drift due to external field is $f$ on the backbone, it is
$-f\cos\alpha$ on the pocket. The geometry is not strictly
one-dimensional, but it is piecewise one-dimensional, as it is
composed by linear segments joined at discrete points. therefore, we
can use the general equations  (\ref{eq:difeq1dforlambda}) on each
segment separately and then guarantee appropriate gluing at the joints,
by imposing proper boundary conditions at the ends of each segment.

The fact that no current flows through the pockets makes the equations
(\ref{eq:difeq1dforlambda}) particularly simple inside the pockets, namely
\begin{equation}
\begin{split}
&D_{0s}
\frac{\partial\lambda_s(x)}{\lambda_s(x)\partial
x}
=-f_{0s}\cos\alpha\\
&D_{0b}
\frac{\partial\lambda_b(x)}{\lambda_b(x)\partial
x}=
-f_{0b}\cos\alpha\;.
\label{eq:difeq1dforlambdajzero}
\end{split}
\end{equation}
We denoted $f_{0s}$ and  $f_{0b}$ the drift at infinite dilution on
the backbone for small
and big particles, respectively, and we have taken into account
tilting of the pockets as shown n Fig.  \ref{scheme-pockets}. The
coordinate $x$ parametrizes the position on the pocket, starting with
$x=0$ at the joint with the backbone and increasing toward $x=W$ at the end of
the pocket.

Technically, the most important point is that the fugacities for small
and big particles decouple, so each equation of the pair
(\ref{eq:difeq1dforlambdajzero}) can be solved separately. The
exponential dependence, or barometric formula, which holds for
densities in the case of non-interacting particles, holds for
fugacities when interaction is taken into account by our hydrodynamic
approximation. Therefore
\begin{equation}
\begin{split}
&\lambda_s(x)=\lambda_{s0}\mathrm{e}^{-xf_{0s}\cos\alpha/D_{0s}}\\
&\lambda_b(x)=\lambda_{b0}\mathrm{e}^{-xf_{0b}\cos\alpha/D_{0b}}\;.
\label{eq:barometricforlambda}
\end{split}
\end{equation}
The average densities and currents (which flow along the backbone)
depend on just two parameters, namely the fugacities
$\lambda_{s0}\equiv\lambda_s(0)$ and
$\lambda_{b0}\equiv\lambda_b(0)$. Of course, the densities, and
therefore also the fugacities, are constant throughout the backbone
and these uniform values are equal to the boundary value at $x=0$ for
the pockets. We obtain
\begin{equation}
\begin{split}
&\overline{\rho}_s=\frac{LR_s(\lambda_{s0},\lambda_{b0})+
\int_0^WR_s(\lambda_{s}(x),
\lambda_{b}(x))dx}{L+W}
\\
&\overline{\rho}_b= \frac{LR_b(\lambda_{s0},\lambda_{b0})+
\int_0^WR_b(\lambda_{s}(x), \lambda_{b}(x))dx}{L+W}
\label{eq:pocketaverdens}
\end{split}
\end{equation}
for the average densities of small and big particles, respectively,
where the dependence of fugacities on the coordinate $x$ along the
pocket is given by (\ref{eq:barometricforlambda}). For the currents,
we obtain simple formulas
\begin{equation}
\begin{split}
&j_s=f_{0s}\frac{R_s^2(\lambda_{s0},\lambda_{b0})}{\lambda_{s0}} \\
&j_b=f_{0b}\frac{R_b^2(\lambda_{s0},\lambda_{b0})}{\lambda_{s0}}\;.
\label{eq:pocketcurr}
\end{split}
\end{equation}

\begin{figure}[t]
\includegraphics[scale=0.85]{%
\slaninafigdir/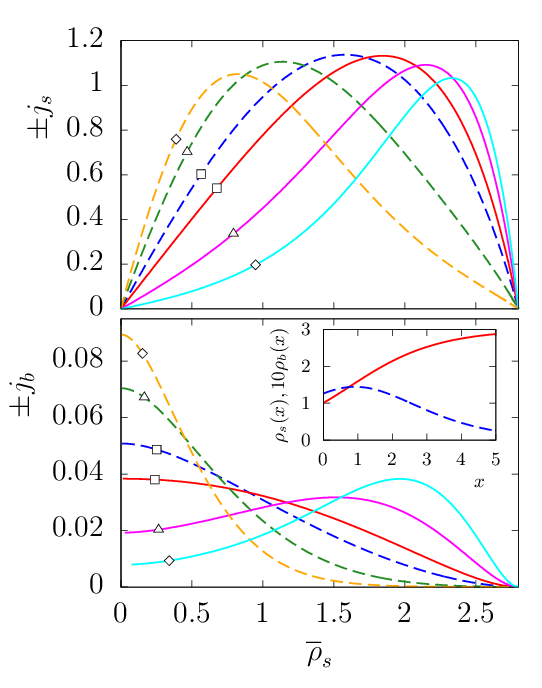}
\caption{Dependence of the current of small (upper panel) and big
(lower panel) particles on the
average density of small particles, for average density of big
particles $\overline{\rho}_b=0.1$. The transport parameters are
$f_0=1$, $D_0=1$ and geometry parameters $L=1$, $\cos\alpha=0.7$.
The dashed lines correspond to
the ``$+$'' sign in front of $j_s$ and $j_b$ (i.e. positive current),
solid lines correspond to   the ``$-$'' sign in front of $j_s$ and
$j_b$ (i.e. negative current). The lines are marked by symbols which
correspond to parameters $W=1$ ($\Box$), $W=3$ ($\bigtriangleup$),
and $W=5$ ({\Large $\diamond$}). In the inset in the lower panel,
local density of small (solid line) and big (dashed line) particles
within the pocket. The
transport parameters are  $f_0=-1$, $D_0=1$,  the geometry parameters
are $L=1$, $W=5$, $\cos\alpha=0.7$, and the average densities
$\overline{\rho}_s=2$ and $\overline{\rho}_b=0.1$.
}
\label{jsjb-pockets-jb0p1-cos0p7-l1-wx}
\end{figure}

\begin{figure}[t]
\includegraphics[scale=0.85]{%
\slaninafigdir/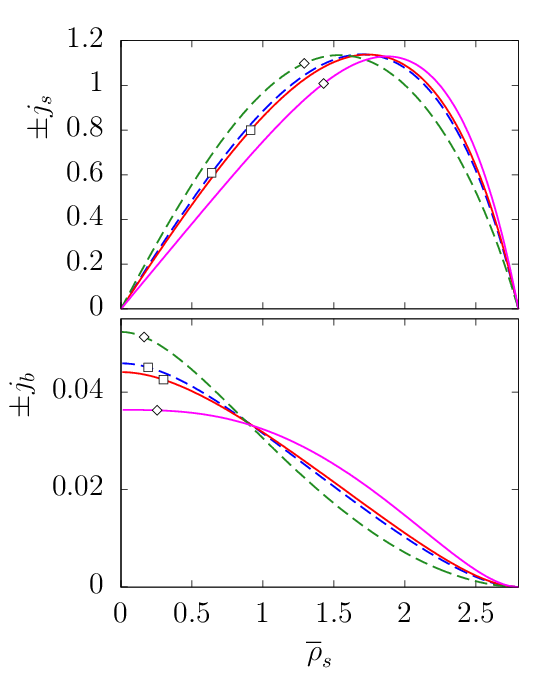}
\caption{Dependence of the current of small (upper panel) and big
(lower panel) particles on the
average density of small particles, for average density of big
particles $\overline{\rho}_b=0.1$.
The transport parameters are
$f_0=1$, $D_0=1$ and geometry parameters $L=1$, $W=1$..
The dashed lines correspond to
the ``$+$'' sign in front of $j_s$ and $j_b$ (i.e. positive current),
solid lines correspond to   the ``$-$'' sign in front of $j_s$ and
$j_b$ (i.e. negative current). The lines are marked by symbols which
correspond to parameters  $\cos\alpha=0.1$ ($\Box$)
and  $\cos\alpha=0.9$ ({\Large $\diamond$}).
}
\label{jsjb-pockets-jb0p1-cosx-l1-w1}
\end{figure}

We focused on the specific case $k=3$, where
\begin{equation}
\begin{split}
&R_s(\lambda_{s},\lambda_{b})=\frac{\lambda_s+\lambda_s^2+\frac{1}{2}\lambda_s^3+\lambda_s\lambda_b
}{1+\lambda_s+\frac{1}{2}\lambda_s^2+\frac{1}{6}\lambda_s^3+\lambda_b+\lambda_s\lambda_b}\\
&R_b(\lambda_{s},\lambda_{b})=\frac{\lambda_b+\lambda_s\lambda_b
}{1+\lambda_s+\frac{1}{2}\lambda_s^2+\frac{1}{6}\lambda_s^3+\lambda_b+\lambda_s\lambda_b}\;.
\label{eq:k3tworfunctions}
\end{split}
\end{equation}
Especially, we are interested in how the geometry, i.e. the
length and angle of inclination of the pockets, influences the
stationary current along the backbone.  In
Figs. \ref{jsjb-pockets-jb0p1-cos0p7-l1-wx} and
\ref{jsjb-pockets-jb0p1-cosx-l1-w1}  we show
the dependence of the currents of small and big particles on the
average density of small particles, with average density of big
particles fixed. Having in mind the ratchet effect, we plot the
results for both orientations of the driving, i.e. for $f_0=|f_0|$
and for $f_0=-|f_0|$. In all these results we assume that both diffusion
coefficient and drift of big particles are half of those of small
particles. This way we take into account the fact that transport
coefficients are inversely proportional to particle size. Therefore,
$D_{0s}=D_0$, $D_{0b}=D_0/2$, $f_{0s}=f_0$, and $f_{0b}=f_0/2$.

\begin{figure}[t]
\includegraphics[scale=0.85]{%
\slaninafigdir/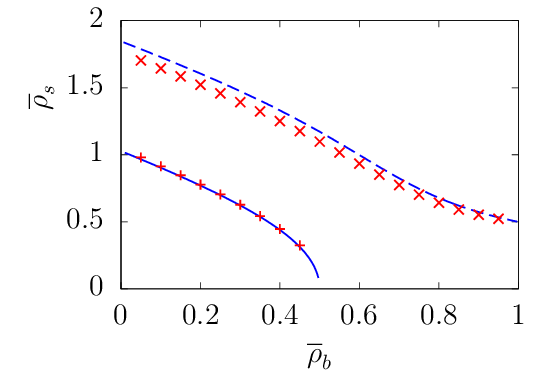}
\caption{Phase diagram of the ratchet currents of small and big
particles in the pocket geometry with parameters $L=1$, $W=1$
$\cos\alpha=0.7$, $D_0=1$ .
{The region of full separation, i. e. when
the ratchet current of small particles is positive, while the
ratchet current of big particles is negative, is delimited by a pair
of lines $\overline{\rho}_{sc2}(\overline{\rho}_{b})$ and
$\overline{\rho}_{sc1}(\overline{\rho}_{b})$.
Two pairs of lines are shown for two different forces
$f_0$. For $f_0=1$ it is the pair of solid line (from below) and
dashed line (from above). For $f_0=5$ the shapes of the lines in the
pair are
indicated by
symbols. The symbols $\times$ indicate the line delimiting the
region from above, the symbols $+$ indicate the line delimiting the
region from below. Note that the lower lines nearly coincide for
$f_0=5$ and $f_0=1$, while the upper lines slightly differ for this
two values of $f_0$. }
}
\label{pockets-phasediag-cos0p7-l1-w1}
\end{figure}

In Fig. \ref{jsjb-pockets-jb0p1-cos0p7-l1-wx} we compare results for several lengths
$W$ of the pocket. Let us first look at the current of small
particles. As expected, the ``easy'' direction is positive,
i.e. rightward, as long as $\cos\alpha >0$. This means that for
small $\overline{\rho}_s$ the absolute value  of the current is larger
for $f_0=|f_0|$ than for  $f_0=-|f_0|$.  However, contrary to the
case of sawtooth potential investigated in the previous section, in
the pocket geometry this holds only for densities
$\overline{\rho}_s<\overline{\rho}_{sc1}$. At the critical density
$\overline{\rho}_{sc1}$ the ratchet current of small particles changes
sign. This can be described by saying that the ``easy'' direction for
holes in opposite to the ``easy'' direction for particles.
Similar picture holds also for big particles. The ratchet current is
positive for $\overline{\rho}_s<\overline{\rho}_{sc2}$ and negative
beyond the second critical density $\overline{\rho}_{sc2}$. For the
data shown in Fig. \ref{jsjb-pockets-jb0p1-cos0p7-l1-wx} we observe
that $\overline{\rho}_{sc2}<\overline{\rho}_{sc1}$. Therefore, there
is an interval of densities
$\overline{\rho}_{s}\in(\overline{\rho}_{sc2},\overline{\rho}_{sc1})$
where ratchet current of small particles is positive, while ratchet
current of big particles is negative. The
Fig. \ref{jsjb-pockets-jb0p1-cos0p7-l1-wx} shows the situation for
just one fixed value of the density of big particles
$\overline{\rho}_{b}$. For other values of $\overline{\rho}_{b}$ the
critical densities $\overline{\rho}_{sc2}$ and
$\overline{\rho}_{sc1}$ may change or even disappear. We can display
the situation by a phase diagram in the axes  $\overline{\rho}_{s}$
and $\overline{\rho}_{b}$. The two critical densities, as functions of
$\overline{\rho}_{b}$, i. e.
$\overline{\rho}_{sc2}(\overline{\rho}_{b})$ and
$\overline{\rho}_{sc1}(\overline{\rho}_{b})$, define lines which
delimit the region of densities for which the ratchet currents of
small and big particles have opposite sign (we shall call it region of
full separation). {In this region we have
$\overline{\rho}_{sc2}(\overline{\rho}_{b})<\overline{\rho}_{s}
<\overline{\rho}_{sc1}(\overline{\rho}_{b})$.}
We show the phase diagram in
Fig. \ref{pockets-phasediag-cos0p7-l1-w1} for two sets of
parameters. As we can see, the region of full separation forms a
diagonal band extending from big $\overline{\rho}_{s}$ and small
$\overline{\rho}_{b}$ to big $\overline{\rho}_{b}$ and small
$\overline{\rho}_{s}$. If we imagine a device intended for separation
of big particles from small particles in a mixture using ratchet
effect in our pocket
geometry, it indeed separates the two types of particles perfectly,
as long as the densities remain in the region of full separation. The
diagonal shape of the region means, that during the separation process
the densities of small and big particles may in principle evolve so
that they remain all the time in the region of full separation and at
the end all big
particles are accumulated at one end and all small particles at the
opposite end of the device. Of course, practical application of such
scenario would require significant amount of fine tuning of the
apparatus.

In Fig. \ref{pockets-phasediag-cos0p7-l1-w1} we can also see that
increasing the driving force from $f_0=1$ to  $f_0=5$ has only very
small influence on the shape of the  region of full separation. If we
look back to  Fig. \ref{jsjb-pockets-jb0p1-cos0p7-l1-wx}, we
observe that the critical densities $\overline{\rho}_{sc2}$ and
$\overline{\rho}_{sc1}$ depend very little on the geometric parameter
$W$ (the length of the pocket). The shape of the curves $j_s$ and
$j_b$ as functions of  $\overline{\rho}_{s}$ varies significantly when
we increase $W$, but the density at which the two curves
corresponding to positive and negative $f_0$ cross, remains nearly the
same. Similar conclusion can be deduced from
Fig. \ref{jsjb-pockets-jb0p1-cosx-l1-w1}, where the current is shown
for several inclinations $\alpha$ of the pocket. Again, the current itself
does depend on the angle $\alpha$, but the critical densities remain
nearly unchanged. All these observations imply that the shape of the
region of full separations, as shown in
Fig. \ref{pockets-phasediag-cos0p7-l1-w1} is very robust and nearly
independent of the details of the geometry and of the strength of the
driving $f_0$. Let us also note that a phase diagram very similar to
Fig.  \ref{pockets-phasediag-cos0p7-l1-w1} was obtained in our
previous work \cite{hum_kot_net_sla_20} where we studied a discrete
model by direct numerical simulations.


Let us turn once more to
Fig. \ref{jsjb-pockets-jb0p1-cos0p7-l1-wx}. For $W=1$ the dependence
of $j_s$ on $\overline{\rho}_{s}$ is concave, but for $W=3$ inflection
point appear and this is even more pronounced for $W=5$. It
corresponds to enhancement of the particle current over the linear
dependence. We observed this phenomenon on the previous section in the
model with sawtooth potential (see
Fig. \ref{j-1d-k2-D0p1-v0p5-lpm0p3and0p7-f0p3}). The explanation is
again the same. Indeed, repulsive interaction between particles
prevents them from filling the pockets beyond the maximum density,
therefore more particles remain on the backbone and these extra particles
are responsible for the interaction-induced enhancement of the
current. A bit more subtle is the effect observed in the dependence of
the current of big particles on  $\overline{\rho}_{s}$. For negative,
i.e. leftward driving, $f_0<0$, and for $W=3$ and $W=5$, we observe
that the current of big
particles  increases when density of small particles increases. This
seems counterintuitive, as one would guess that small particles act as
obstacles for the movement of big ones. However, the actual scenario
is more tricky. The small
particles quickly fill the pockets thus forcing the big
particles to remain in the backbone. This is illustrated in the inset
in Fig.  \ref{jsjb-pockets-jb0p1-cos0p7-l1-wx}. We can see how the
densities of small and big particles depend on the position inside the
pocket. The driving is $f_0=-1$, i.e. the particles are pushed into
the pocket. While the density of small particles increases
monotonously when we proceed from the entrance to the end of the
pocket, the big particles behave differently. Initially, their density
increases a little, but when we proceed deep inside the pocket, the
density of big particles decreases substantially. Small particles push
the big ones out of the pocket, countering and reversing  the effect
of the driving force.
As a result, the current of big
particles is enhanced. Note that this is analogous to the
Brazil nut phenomenon, occurring frequently in shaken mixtures of
granular matter \cite{ro_stra_pri_swe_87}. We shall encounter this
effect at several occasions later.

\section{Two-dimensional channels}

\subsection{Geometry}

Let us now turn to more realistic geometries. In this Section we shall
consider channels with periodically varying profile in the horizontal
direction, but with uniform height in the vertical direction,
comparable with the size of the particles. Therefore, the movement of
the particles can be considered effectively two-dimensional. Such
structures are routinely fabricated using PDMS soft lithography
\cite{mar_bug_tal_sil_02,whitesides_06} and this is the experimental
situation we have in mind in our theoretical modeling.

\begin{figure}[t]
\includegraphics[scale=0.3]{%
\slaninafigdir/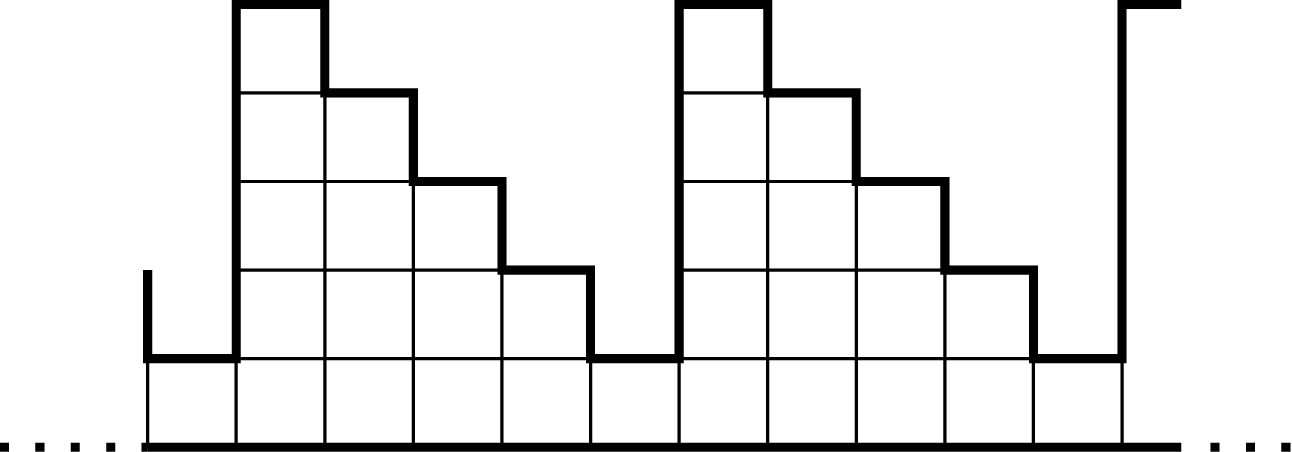}
\caption{Scheme of the two-dimensional channel geometry. The channel
is composed of periodically repeating teeth, each tooth consisting of
$15$ squares drawn by the thin lines. The edges of the squares have unit length. In the
continuous description, the position of the particles can be
anywhere within the area delimited by the borders shown as the thick
lines. In the discrete description, the number of particles within
each square must satisfy the condition (\ref{eq:genasepzuby}).
}
\label{scheme-zuby}
\end{figure}

We shall investigate the dynamics of particles in 2D channel with
geometry sketched in Fig. \ref{scheme-zuby}. The channel consists of a
series of identical teeth. Each tooth is composed of $15$ equal
square cells. The edge of a single square sets the unit of length.

\begin{figure}[t]
\includegraphics[scale=0.85]{%
\slaninafigdir/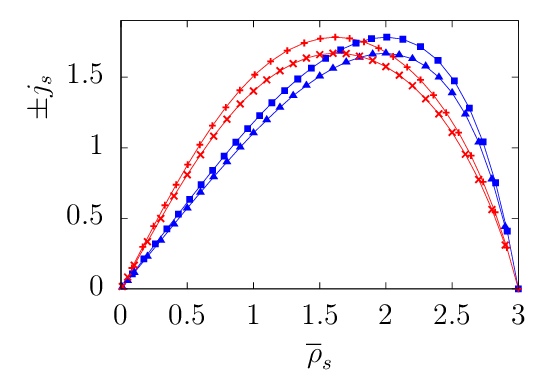}
\caption{Particle current in the one-component system (small particles
only) in the two-dimensional teeth geometry shown in
Fig. \ref{scheme-zuby}.
The diffusion coefficient is $D_0=1$ and the driving $f_0=1$
(symbols $+$ and $\times$ and
sign ``$+$'' in front of $j_s$) and $f_0=-1$ (symbols $\blacksquare$
and $\blacktriangle$ and sign ``$-$'' in front of $j_s$). The points
show the results of the Monte-Carlo simulation of the discrete model
(symbols $\times$ and
$\blacktriangle$) and solution of the equation (\ref{eq:stactwotypes}) using
COMSOL software in the continuous formulation (symbols $+$ and $\blacksquare$).
}
\label{js-k3-rb0}
\end{figure}

\subsection{Discrete model}

In the discrete formulation
we performed Monte-Carlo simulations of the generalized ASEP model in
the geometry shown in Fig. \ref{scheme-zuby}. The dynamical variables
are the numbers of particles in each of the square cells which must
obey the exclusion constraint (\ref{eq:stericdiscrete}). Specifically,
each square cell has
capacity $k=3$, there are just two types of particles (small and
big). The small particles have size factor $c_s=1$, the big ones
have $c_b=2$. Therefore, in each cell the number of small particles
$n_s$ and big particles $n_b$ must satisfy the constraint
\begin{equation}
n_s+2n_b\le k\;.
\label{eq:genasepzuby}
\end{equation}
The sets of hopping rates are, in the direction right, left, up, down,
respectively, denoted $a$, $b$, $c$, $d$ for small particles and $A$,
$B$, $C$, $D$ for big particles. We always fix $c=d$, $C=D$. To keep
compatibility with continuous
description, the free diffusion coefficients and drifts are
$D_{0s}=(a+b)/2=(c+d)/2$, $f_{0s}=a-b$ for small particles and
$D_{0b}=(A+B)/2=(C+D)/2$, $f_{0s}=A-B$ for big particles.
Moreover, as the properties of Brownian motion imply that the
diffusion coefficient and mobility are both inversely proportional to
particle radius, we fix the relation between the hopping rates of
small and big particles as
$A=a/2$, $B=b/2$, $C=c/2$ and $D=d/2$. Therefore, there are just two
free parameters of the model, namely $D_0\equiv D_{0s}$ and
$f_0\equiv f_{0s}$.
The average density of particles of type $\alpha$ is
$\overline{\rho}_\alpha=N_\alpha/(15 L)$, where $L$
is the number of teeth in the sample, each comprising $15$ cells,
and $N_\alpha$ the number of particles of type $\alpha$ in the simulation.

\subsection{Continuous formulation in hydrodynamic approximation}
In the continuous formulation,
we solved numerically the set of equations (\ref{eq:stactwotypes}) in
stationary state in the channel delimited by thick lines in the scheme
shown in Fig. \ref{scheme-zuby}. We fix the $x$-axis along the channel
and $y$-axis perpendicular to it. Consistently with the discrete
formulation, there are just two types of
particles, and we consider the simplest
non-trivial case $k=3$. The functions
$R_\alpha(\lambda_s,\lambda_b)$ are given by the formulas
(\ref{eq:k3tworfunctions}). To keep the transport coefficients
consistent with the discrete model, we fix the diffusion coefficients
$D_{0s}=2D_{0b}=D_0$ and drift $f_{0s}=2f_{0b}=f_0$.
The boundaries (thick lines) impose
reflecting boundary conditions. The teeth repeat periodically, so
that we impose periodic conditions on the fugacities
$\lambda_\alpha(x+5,y)=\lambda_\alpha(x,y)$, as each tooth has length $5$
in the $x$-direction.
The numerical solution is performed using the COMSOL software. The
average density of particles of type $\alpha$ is just the integral of
the local density over one tooth divided by
the area of the tooth,
$\overline{\rho}_\alpha=\int_{\mathrm{tooth}}\rho_\alpha(x,y)dxdy/15$.

\subsection{Comparison}

First, we studied the one-component system with small particles
only. We show in Fig. \ref{js-k3-rb0} the current of particles as a
function of average density.  As expected, we observe the maximum
of the current, which lies at different position depending on the sign
of the drift $f_0$. When the orientation of the drift is switched
periodically, a ratchet effect occurs.
If the frequency of switching is very small, we may use the adiabatic
approximation and calculate the ratchet current as arithmetic mean of
the currents for $f_0=|f_0|$ and $f_0=-|f_0|$. The ratchet current
calculated in such a way can be grasped from  Fig. \ref{js-k3-rb0}
directly. We can see that in the geometry of Fig. \ref{scheme-zuby}
the ratchet current is positive for small densities, corresponding to
the easy direction toward the right. At a density around
$\overline{\rho}_s\simeq 1.8$ the ratchet current changes sign and
becomes negative. Recall that the change of sign of the ratchet
current is absent in the
one-dimensional model investigated in the last Section, but occurs in
the pocket geometry. This indicates that the pocket geometry indeed
grasps essential features of the 2D systems.

Comparing the results of discrete and continuous approaches, we can
see that the hydrodynamic approximation slightly
overestimates the current, but the difference remains rather
small. Moreover, it is interesting to note that the density at which
the ratchet current changes sign is nearly equal for the discrete and
continuous formulations. All of this suggests that the hydrodynamic
approximation is fairly reliable for one-component colloid
suspensions.

\begin{figure}[t]
\includegraphics[scale=0.85]{%
\slaninafigdir/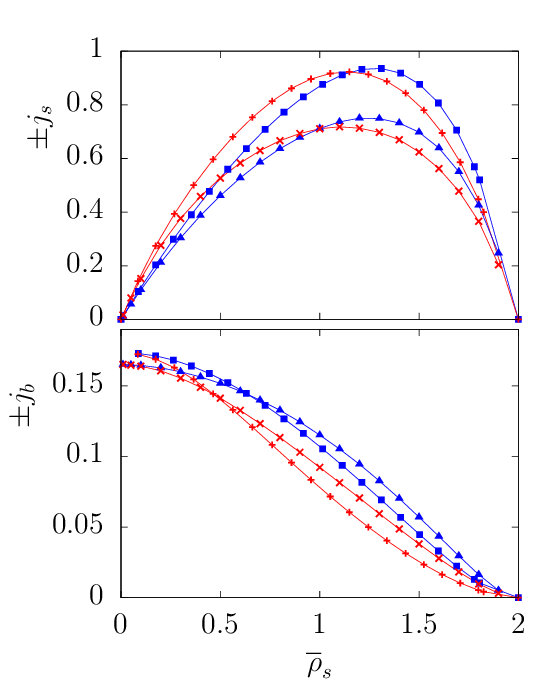}
\caption{Current of the small (upper panel) and big (lower panel) particles
in the two-dimensional teeth geometry shown in
Fig. \ref{scheme-zuby}.
The diffusion coefficient is $D_0=1$ and the driving $f_0=1$
(symbols $+$ and $\times$ and
sign ``$+$'' in front of $j_s$) and $f_0=-1$ (symbols $\blacksquare$
and $\blacktriangle$ and sign ``$-$'' in front of $j_s$). The
average density of big particles is $\overline{\rho}_b=0.5$.
The points
show the results of the Monte-Carlo simulation of the discrete model
(symbols $\times$ and
$\blacktriangle$) and solution of the set of equations (\ref{eq:stactwotypes}) using
COMSOL software in the continuous formulation (symbols $+$ and $\blacksquare$).
}
\label{jsjb-k3-rb0p5}
\end{figure}

Next we simulated the mixed system of small and big particles. We show
in Fig. \ref{jsjb-k3-rb0p5} how the current of small and big particles
depend on the  average density of small particles, with average density of big
particles fixed. As expected, we can observe the maximum in the
current of small particles and monotonous decrease in the current of
big particles. Similarly to one-component system, comparison of the
results for positive and negative drift tells us what is the ratchet
current like in adiabatic approximation. Again, we observe current
reversal in the ratchet current of small particles. The current
reversal for big particles is absent in  Fig. \ref{jsjb-k3-rb0p5}, but
this is a special feature of the particular choice of the density of
big particles $\overline{\rho}_b=0.5$. For a generic
$\overline{\rho}_b$, also the current of big particles exhibits the
change of sign.  Thus, also the mixed system of small and big
particles is qualitatively very close when we compare the
piecewise-one-dimensional pocket geometry and the two-dimensional
tooth geometry.

Comparing the results for the discrete and continuous models, we can
see that the agreement is significantly worse than in the case of
one-component system, although qualitatively the behavior remains
comparable. Such observation is consistent with the results we
obtained recently \cite{sla_kot_net_22} on one-dimensional
generalized ASEP model. Indeed, in  \cite{sla_kot_net_22} we showed
that the mean-field and Kirkwood approximations are excellent for
one-component system, but become quantitatively off by several tens
of per cent if we work with two-component system. In
\cite{sla_kot_net_22} we identified the source of the disagreement to
be long-range correlations which develop due to indirect interaction
of one type of particles mediated by the other type of particles. We
believe that the same mechanism is responsible for the difference
between discrete and continuous models also here.

\begin{figure*}[t]
\includegraphics[scale=0.75]{%
\slaninafigdir/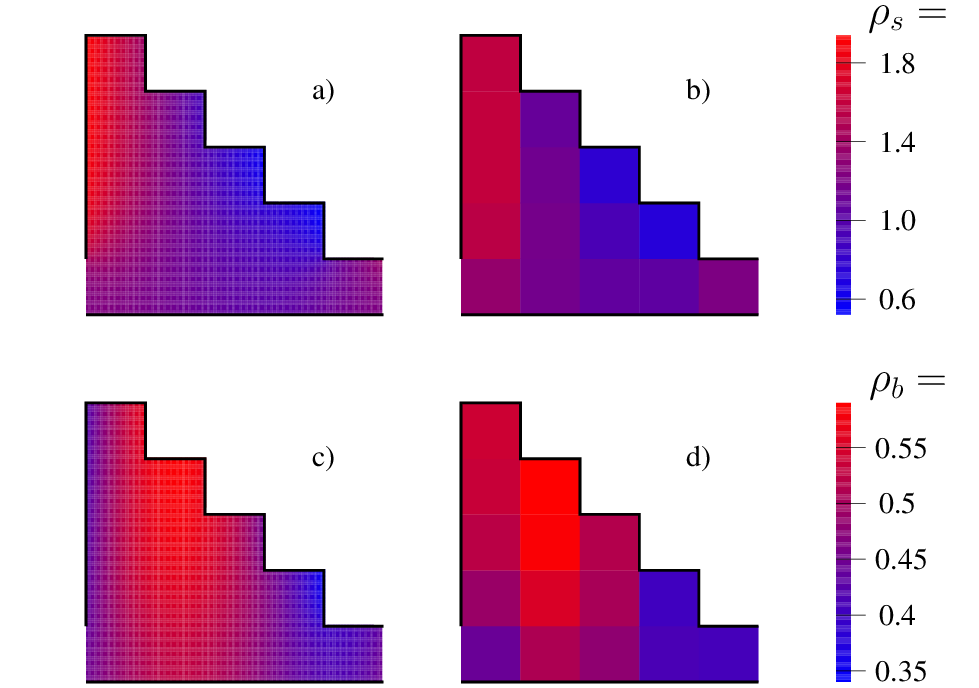}
\caption{Local densities of small (panels a) and b)) and big (panels
c) and d)) particles, obtained from continuous formulation (panels
a) and c)) and from discrete model (panels b) and d)) in the
two-dimensional teeth geometry shown in
Fig. \ref{scheme-zuby}.  The transport
coefficients are $D_0=1$, $f_0=-1$. The average density of small
particles is $\overline{\rho}_s=1.2$ and the average density of big
particles is  $\overline{\rho}_b=0.5$. The color encoding for the
local densities  is given by the legends on the right-hand
side.
}
\label{k-asep-zuby-simul}
\end{figure*}

In Fig. \ref{k-asep-zuby-simul} we can see the spatial distribution
of small and big particles within one tooth. The orientation of
the drift is leftward and this corresponds to marked accumulation of
small particles at the left edge of the tooth. On the contrary,
big particles are localized mostly near the middle of the
tooth. This is another manifestation of the phenomenon analogous
to Brazil nut effect. Indeed, small particles accumulated next to the
left edge of the tooth prevent the big particles from entering the
leftmost region. The big particles, which are also a priori pushed
leftward, remain at a half way. This phenomenon can be seen in both
discrete and continuous models. Also other details of the local
density are very similar in discrete and continuous variants, so we can
conclude that the continuous description works well in terms of
spatial distribution of the particles.


\section{three-dimensional geometries}

\subsection{Mixture of particles in a container:
The Brazil nut effect}

In static case, i.e. with strictly zero current everywhere, the
equations for fugacities (\ref{eq:stactwotypes}) decouple. For
example, we consider a mixture of two types of particles in a
cylindric container, with vertical axis, under influence of gravity. We identify the
vertical direction with the $z$-axis and we have homogeneous field
$f_{0x}=f_{0y}=0$, $f_{0z}=-f_0$. The solution of
(\ref{eq:stactwotypes}) is trivial
\begin{equation}
\begin{split}
\lambda_s(x,y,z)=\lambda_{s_0}\mathrm{e}^{-\frac{f_0}{D_0}z}\\
\lambda_b(x,y,z)=\lambda_{b_0}\mathrm{e}^{-\frac{f_0}{D_0}z}
\end{split}
\label{eq:3dcontainerlambdas}
\end{equation}
and the densities of particles are deduced directly from
(\ref{eq:3dcontainerlambdas}) using the functions
(\ref{eq:k3tworfunctions}) as
\begin{equation}
\begin{split}
\rho_s(x,y,z)=R_s(\lambda_{s_0}\mathrm{e}^{-\frac{f_0}{D_0}z},\lambda_{b_0}\mathrm{e}^{-\frac{f_0}{D_0}z})\\
\rho_b(x,y,z)=R_b(\lambda_{s_0}\mathrm{e}^{-\frac{f_0}{D_0}z},\lambda_{b_0}\mathrm{e}^{-\frac{f_0}{D_0}z})\;.
\end{split}
\label{eq:3dcontainerrhos}
\end{equation}

\begin{figure}[t]
\includegraphics[scale=0.85]{%
\slaninafigdir/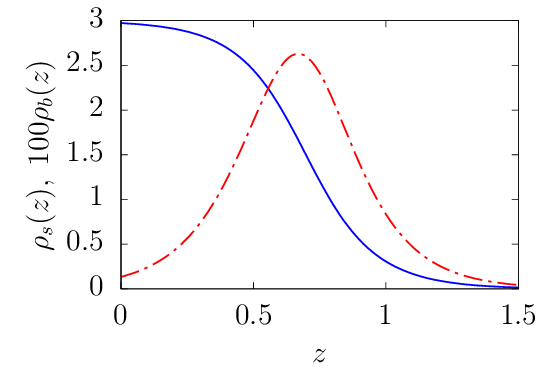}
\caption{The Brazil nut effect: local densities of small (solid
line) and big (dot-dashed line) particles in a cylinder of height
$h=1.5$ with vertical axis along the $z$-coordinate. The average
densities are $\overline{\rho}_s=1.4$ for small particles and
$\overline{\rho}_b=0.01$ for big particles. The transport
coefficients are $D_0=1$, $f_0=6$.
}
\label{rsrb-3d-brazilnut-rs1p4-rb0p01-h1p5-f6}
\end{figure}

We show in  Fig. \ref{rsrb-3d-brazilnut-rs1p4-rb0p01-h1p5-f6} the
typical density profile of small and big particles.  In this example
we assume that there are just few big particles, i. e. the average
concentration of big particles is much smaller than the concentration
of small ones. The small
particles accumulate at the bottom, with diffuse but steep drop of
density around a depth analogous to a ``liquid level''. Then, the big
particles accumulate around this ``liquid level'', thus resembling big
particles floating on the top of the bulk of small particles. This is
just what is observed in the Brazil nut phenomenon. However, there is
a fundamental difference. In the Brazil nut experiments \cite{ro_stra_pri_swe_87}, one uses
shaken granular material, while in our model, we deal with Brownian
particles under the influence of thermal fluctuations.

In this formulation, the Brazil nut effect occurs in a static
regime with no macroscopic current. As we have seen in the last two
sections, it can be observed also in non-equilibrium situations. So,
we can conjecture that it is a generic feature of mixed systems of
small and big particles.

\begin{figure}[t]
\includegraphics[scale=0.4]{%
\slaninafigdir/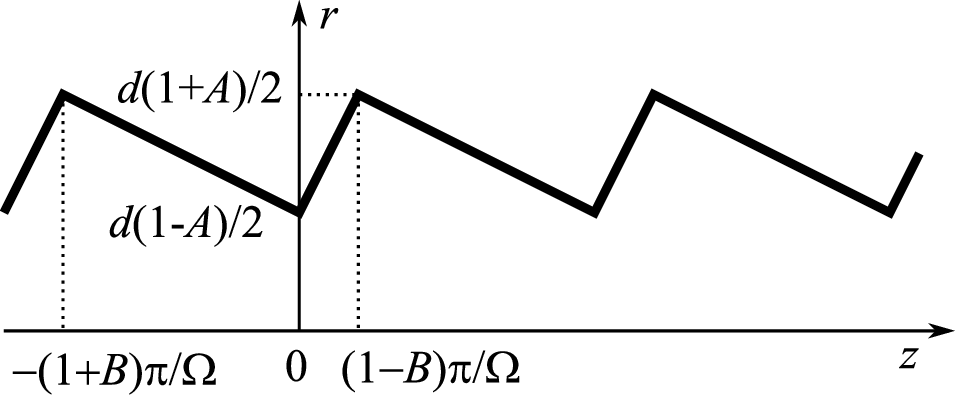}
\caption{The profile of the axially symmetric pore. The $z$-axis
coincides with the axis of the pore. The geometry is characterized
by parameters $d$ (average diameter), $\Omega$ (spatial frequency),
$A$ (depth of the profile) and $B$ (asymmetry of the profile).
}
\label{scheme-3daxisym}
\end{figure}

\subsection{Axially symmetric pore}

Now we consider the mixture of small and big particles inside an
axially symmetric pore, with driving parallel to the axis.
We use cylindrical coordinates with axis $z$
coinciding with the axis of the pore and radial coordinate $r$. All
quantities will be symmetric with respect to rotation around the axis,
so the dependence on the azimuthal angle $\phi$ is absent.
The pore is delimited by a periodic wall, $r<h(z)$,
$h(z)=h(z+2\pi/\Omega)$. For simplicity, we shall use a sawtooth-type
boundary, as sketched in Fig. \ref{scheme-3daxisym}. There are three
geometric parameters, $d$ specifying the average diameter of the pore,
$\Omega$ the spatial frequency,
$A$ depth of the profile and $B$ asymmetry of the profile.
Specifically, we
have
\begin{equation}
\begin{split}
h(z)= \frac{d}{2}&\Big(1+A(\frac{2z}{z_1}-1)\Big),\;z\in(0,z_1)\\
h(z)= \frac{d}{2}&\Big(1-A(\frac{2z}{z_2}+1)\Big),\;z\in(-z_2,0)
\end{split}
\label{eq:3daxisymprofile}
\end{equation}
where we denoted $z_1=(1-B)\pi/\Omega$ and  $z_2=(1+B)\pi/\Omega$.

\begin{figure}[t]
\includegraphics[scale=0.85]{%
\slaninafigdir/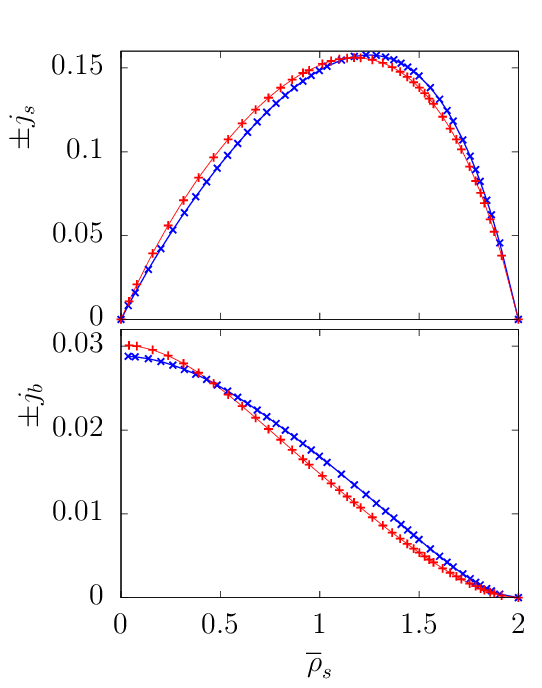}
\caption{Current of small (upper panel) and big (lower panel)
particles in axially symmetric pore with geometry defined by
(\ref{eq:3daxisymprofile}).  The geometric parameters are $d=0.3$, $\Omega=20$, $A=0.3$,
$B=0.8$, the transport coefficients are $D_0=0.2$, $f_0=5$ (symbols
$+$,  sign ``$+$'' in front of $j_a$ and $j_b$)
and  $f_0=-5$ (symbols $\times$, sign ``$-$'' in front of $j_a$ and $j_b$). The
average density of big particles is  $\overline{\rho}_s=0.5$.
}
\label{jsjb-k3-3daxisym-rb0p5}
\end{figure}

We solved the transport equations (\ref{eq:stactwotypes}) using the
COMSOL software. From a technical point of view, the axial symmetry
facilitated substantially the numerical solution. In
Fig. \ref{jsjb-k3-3daxisym-rb0p5} we show the current of small and big
particles as a function of the average density of small
particles. Comparing the results for rightward and leftward driving,
we can immediately infer the properties of the ratchet current in
adiabatic approximation (i.e. when periodic flipping of the
orientation of the drift occurs very slowly). We can see that the
ratchet current of both small and big particles is positive if the
density of small particles is small enough. This corresponds to moving
in the easy direction dictated by the geometry sketched in
Fig. \ref{scheme-3daxisym}. Indeed, the easy direction is rightward as
long as the asymmetry parameter $B>0$, which is the case for the data
in Fig. \ref{jsjb-k3-3daxisym-rb0p5}.
When the density of small
particles increases, the ratchet current of big particles becomes
negative first and the ratchet current of small particles later. This
means that there is an interval of densities $\overline{\rho}_s$ in
which the ratchet effect carries the small and big particles in
opposite directions, thus enabling full separation of the particle
types. This is the same effect as observed already in the piecewise
one-dimensional geometry with pockets, studied in Sec. III.C. Also the
whole current-density diagram of a 3D system, Fig.
\ref{jsjb-k3-3daxisym-rb0p5}, is qualitatively very similar to
corresponding diagram for pocket geometry,
Fig. \ref{jsjb-pockets-jb0p1-cosx-l1-w1}.  This
indicates that the pocket geometry, when properly calibrated, can
bring useful information even on the behavior of 3D systems. We would
like to consider that a late justification of our earlier studies of
generalized ASEP models in pocket geometry
presented in our previous work \cite{hum_kot_net_sla_20}.

\begin{figure}[t]
\includegraphics[scale=0.85]{%
\slaninafigdir/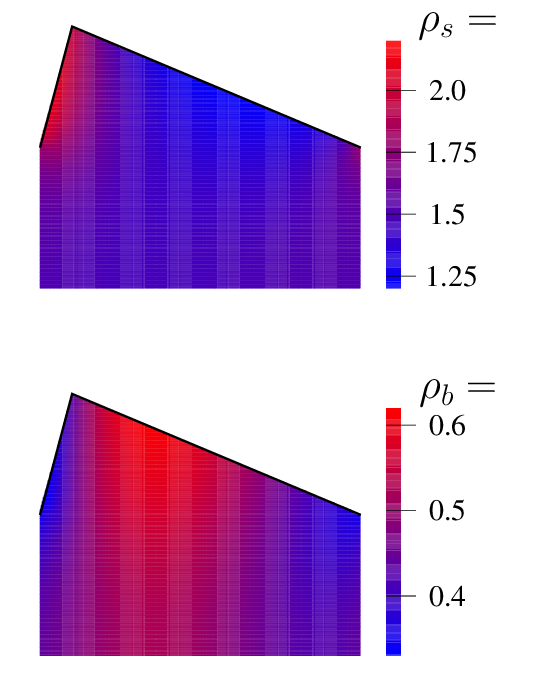}
\caption{Spatial distribution of the particle densities for small
particles (upper panel) and big particles (lower panel) inside one
spatial period of the pore with profile defined by
(\ref{eq:3daxisymprofile}). The color
encoding for local densities is given by the legends on the right-hand
side. The geometric parameters are $d=0.3$, $\Omega=20$, $A=0.3$,
$B=0.8$, the transport coefficients are $D_0=0.2$, $f_0=-5$, and the
average particle densities are $\overline{\rho}_s=1.5$,  $\overline{\rho}_s=0.5$.
}
\label{density-3daxisym-fm5-rs1p5-rb0p5}
\end{figure}

We also looked at the detailed distribution of densities of small and
big particles in the pore. Due to the symmetry, the density depends
only on the axial and radial coordinates. Thus, we show in
Fig. \ref{density-3daxisym-fm5-rs1p5-rb0p5}
the densities in a plane containing the axis of the pore. We also
assume periodicity of the density along the axis and show just one
spatial period of the pore. In the case shown  in
Fig. \ref{density-3daxisym-fm5-rs1p5-rb0p5} the drift goes leftward
and we can see accordingly the accumulation of small particles at the
left wall of the pore. The big particles, on the other hand, are
concentrated around the middle of the period, manifesting once again
the Brazil nut phenomenon. Indeed, the spatial distributions of
particles in 3D geometry in Fig.
\ref{density-3daxisym-fm5-rs1p5-rb0p5} and in 2D geometry in
Fig. \ref{k-asep-zuby-simul} are qualitatively very similar.


\section{Conclusions}

We formulated a hydrodynamic approximation for movement of dense
colloidal suspension of Brownian particles. The formulation goes in
two steps, first approximating the true Brownian motion with steric
repulsion of particles by a discrete stochastic model, the generalized
asymmetric exclusion process (ASEP) and then performing hydrodynamic
limit in this generalized ASEP model. The approximation is formulated
for mixtures of particles of various types, differing by size,
diffusion coefficient and drift. In practical calculation, we worked
with just two types of particles (we called them small particles and
big particles).  An important feature of our
approach is that the transport equations are formulated in terms of
fugacities corresponding to the particle types, rather than in terms
of particle densities. In such formulation, there are situations in
which the equations for several particle types decouple and each
particle type can be treated independently. We investigated several
such situations.

We studied the model in one- two- and three-dimensional geometries. In
a truly 1D system, we placed the driven one-component system (small
particles only)  on
a line under the influence of periodic sawtooth potential. In this
case, the transport equation can be solved explicitly. We observed
several interesting features of the dependence of current on average
density. For small densities, current grows faster than linearly with
density of particles. At certain density the current reaches a maximum
and then decreases when the density approaches its maximum dictated by
the cell capacity. Close to the maximum density, we can look at the
current-density diagram as showing the dependence of current on the
density of holes. Again, we observe that the current increases faster
than linearly with the density of holes. This super-linear behavior at both
low and high densities can be easily interpreted as a consequence of
steric repulsion of particles. Indeed, non-interacting particles
accumulate near the minima of the sawtooth potential. This leads to
suppression of the current, because particles trapped at the minimum
must overcome a
potential barrier.  Steric repulsion puts a  limit to the accumulation
of particles anywhere, therefore less particles are located at the
minima and the spatial distribution of particle density is more uniform.
Hence less particles are slowed down by the potential barrier and the
current may be higher.

We also studied the piecewise one-dimensional case of pocket
geometry. In this setting, linear segments (called pockets)
are appended at regularly
spaced points to a straight line.  In stationary state the current is
constant and non-zero only on the straight line, while it is zero in
the pockets. This is one of the cases when transport equations for
different types of particles decouple and are solved independently. In
this case, the solution of the equations is particularly easy. In
fact, the spatial dependence of fugacities for both small and big
particles follow the barometric formula, familiar from the problem of
free Brownian particles in static gravitational field. From this
observation we deduced the current-density diagram for both types of
particles and density dependence of the ratchet currents in adiabatic
approximation. In this case, the ratchet current changes sign at
certain value of the density. There is a region of densities in which
the ratchet currents of small and big particles are oriented in
opposite direction, therefore enabling (in principle) perfect
separation of the two particle types. In the
two-dimensional diagram with densities of small and big particles on
horizontal and vertical axes, respectively, we can identify the
region where the ratchet currents point in opposite
side. Interestingly, this region forms a strip going diagonally from
pure small particles to pure big particles. This
indicates that we can start with a mixture of small and big particles
and separating gradually the one from the other, we can achieve full
separation, as long as we remain, during the process, all the time
within the strip. Note that similar diagram was obtained by
simulations of generalized ASEP model in our previous work
\cite{hum_kot_net_sla_20}.

{In this work we limited our study of the ratchet
effect to adiabatic approximation. In reality, however, the
direction-switching frequency is finite and represents an important
parameter of the experimental setup. It would be in principle
straightforward to extend our analysis to arbitrary time-dependent
driving by solving the full time-dependent non-linear diffusion
equation (\ref{eq:difeqSB}). However, currently we do not see a way
to obtain analytic results and the only way would be numerical
solution. This could not be complicated using the COMSOL software
but we leave it for future work. We expect that the ratchet current
would be a decreasing function of frequency, ultimately approaching
zero for very fast switching. However, we cannot exclude a
stochastic resonance leading to a maximum current at finite
frequency. This question must be answered by a specific detailed computation.}

In a two-dimensional geometry, we placed the particles in a periodic
tooth-shaped channel. We compared the continuous hydrodynamic
description with simulation of discrete generalized ASEP model in the
same geometry. The results are quantitatively close to each other. In
the case of one-component system (small particles only), the agreement
is very good. However, for mixed small and big particles, the
agreement worsens, as the continuous description cannot take into
account complicated long-range correlations. This is similar to our
observation we made in our recent work on generalized ASEP model
\cite{sla_kot_net_22}.

On the other hand, the continuous hydrodynamic description correctly
describes one important phenomenon involving correlations between
small and big particles, which is the Brazil nut effect. It is
manifested in depletion of big particles from the areas where small
particles concentrate. For example, it implies that the driven small
particles in the tooth-shaped channel accumulate at the posterior
walls of the teeth, as expected, but the big particles are mostly
located around the middle of each tooth, rather than at the posterior
wall. The reason is that the small particles accumulate easily and the
big particles do not find enough space at places occupied already by
small ones. In this respect, the discrete and continuous descriptions
are in full accord. This may be understood well, because the Brazil
nut phenomenon reflects purely local correlations. Such correlations are well
described by the continuous theory, contrary to the long-range
correlations, as shown by us also in the work \cite{sla_kot_net_22}.

We also looked at truly three-dimensional, though simplified,
geometries. We investigated explicitly the Brazil nut effect,
mentioned earlier, in static case of the mixture of small and big
particles in a cylindric container in homogeneous axial external field. The
external field can be viewed as gravitational force. Then, the small
particles are concentrated at the lower part of the container,
resembling a liquid in a half-filled bottle, while
the big particles look as if they floated on top of the surface of the
small-particle liquid. Note, however, that the real experiments with
the Brazil nut phenomenon \cite{ahm_sma_73,
hue_sua_04,sch_ver_kre_swi_swi_06} are made with shaken granular
matter of two
different sizes of grains, while our theory works with Brownian
particles agitated by thermal fluctuations. There is no trivial
correspondence between these two situations, but rather just an
analogy.
{The Brazil nut effect in shaken granular mixtures was amply
studied by molecular dynamics
(e. g. \cite{po_he_95,sae_viv_mel_05,sch_ver_kre_swi_swi_06}) or
schematic Monte Carlo simulations
(e. g. \cite{ro_stra_pri_swe_87}). These studies take into account many
more details of the process than our method, but comparison of
stationary density profiles (e. g. \cite{sae_viv_mel_05}) shows
striking similarity to our results. Therefore, our method grasps
reasonably well
the stationary state. However, there is much more to Brazil nut
effect, for example conduction eddies \cite{po_he_95}, friction and
arching \cite{sae_viv_mel_05}, that our approach cannot explain.
}

The most realistic case we studied in this paper was the mixture of
small and big particles in an axially-symmetric pore with periodically
variable diameter. The variation of the diameter follows a sawtooth
pattern. Therefore, the axial section of the pore resembles the
two-dimensional tooth geometry investigated earlier. Numerical
solution of the transport equations provided concentration dependence
of the current of small and big particles, which is qualitatively very
similar to both the two-dimensional tooth channel geometry and the
one-dimensional pocket geometry. Also the spatial distribution of the
small and big particles in the axial section of the pore is very
similar to what is observed in the two-dimensional system.

This observation provides an a posteriori
justification of the generalized ASEP model in pocket geometry, which
we introduced in \cite{hum_kot_net_sla_20}. Most importantly, the full
three-dimensional solution exhibits qualitatively the same behavior of
the ratchet current. This implies that such a setup can be indeed used
for separation of dense mixtures of colloid particles according to
their size. Quantitatively, however, the ratchet current in
realistic three-dimensional geometry is smaller than in a model
one-dimensional setup, because the wider sections of the pore, which
play the role of pockets seen in one-dimensional model, cannot be made
arbitrarily voluminous, while in one-dimensional case we are not
limited in increasing the depth of the pockets. This implies that we
actually could model the three-dimensional case by one-dimensional pocket
geometry, but with the provision that the depth of the pocket must be
appropriately calibrated and we should expect that the pockets are
relatively shallow.

{
Finally, let us briefly mention a direction for future research. It
is natural, instead of periodically driven Brownian particles, to
consider active particles which possess internal drive and change
direction stochastically
\cite{rom_bar_ebe_lin_12,fod_mar_18}. They were already widely studied in
complex geometries \cite{bec_dil_low_rei_vol_vol_16} and especially
the ratchet effect was explored
\cite{ang_cos_dil_11,gho_mis_mar_nor_13,ai_che_he_li_zho_13,ao_gho_li_sch_han_mar_14,ols_rei_17},
following the experiments with rectification of bacterial movement
(see e. g. \cite{mah_cam_bis_kom_cha_soh_hud_09}). It would be
interesting to use  the
methods developed in our work to dense ensembles of active
particles. We leave this question for future work.
}
\begin{acknowledgments}
We wish to thank K. Neto{\v{c}}n\'y for inspiring discussions and
for providing us with his unpublished results.
\end{acknowledgments}
%
%
%
%
%
%

%
%
%
%
\end{document}